\documentclass[conference]{IEEEtran}
\IEEEoverridecommandlockouts
\usepackage{cite}
\usepackage{amsmath,amssymb,amsfonts}
\usepackage{algorithmic}
\usepackage{graphicx}
\usepackage{textcomp}
\usepackage{xcolor}
\usepackage{url}
\usepackage{color}

\usepackage{soul}
\usepackage{hyperref}
\def\BibTeX{{\rm B\kern-.05em{\sc i\kern-.025em b}\kern-.08em
    T\kern-.1667em\lower.7ex\hbox{E}\kern-.125emX}}
\begin{document}
\title{Efficiency and Optimality in Electrochemical Battery Model Parameter Identification: A Comparative Study of Estimation Techniques
}
\author{\IEEEauthorblockN{1\textsuperscript{st} Feng Guo\thanks{*\textcopyright~2024 IEEE. Personal use of this material is permitted.
Permission from IEEE must be obtained for all other uses, in any current or future media, including reprinting/republishing this material for advertising or promotional purposes, creating new collective works, for resale or redistribution to servers or lists, or reuse of any copyrighted component of this work in other works.
This is the author’s accepted version of the paper published in the \emph{Proceedings of the 2024 10th International Conference on Optimization and Applications (ICOA)}.
The final version is available at IEEE Xplore via DOI:
\protect\href{https://doi.org/10.1109/ICOA62581.2024.10754301}{10.1109/ICOA62581.2024.10754301}.
The implementation of the models used in this paper can be accessed at \protect\url{https://github.com/FrankSuperG/CPG-SPMT}; Zenodo: \protect\url{https://doi.org/10.5281/zenodo.16921607}.}}
\IEEEauthorblockA{\textit{VITO, Mol, Belgium} \\
\textit{EnergyVille, Genk, Belgium}\\
feng.guo@vito.be; \\
0000-0002-5141-8672}
\and
\IEEEauthorblockN{2\textsuperscript{nd} Luis D. Couto }
\IEEEauthorblockA{\textit{VITO, Mol, Belgium} \\
\textit{EnergyVille, Genk, Belgium}\\
luis.coutomendonca@vito.be; \\
0000-0001-8415-7459}
\and
\IEEEauthorblockN{3\textsuperscript{rd} Guillaume Thenaisie}
\IEEEauthorblockA{\textit{VITO, Mol, Belgium} \\
\textit{EnergyVille, Genk, Belgium}\\
guillaume.thenaisie@vito.be; \\
0000-0002-1099-8949}
}

\maketitle

\begin{abstract}
Parameter identification for electrochemical battery models has always been challenging due to the multitude of parameters involved, most of which cannot be directly measured. This paper evaluates the efficiency and optimality of three widely-used parameter identification methods for electrochemical battery models: Least Squares Method (LS), Particle Swarm Optimization (PSO), and Genetic Algorithm (GA). Therefore, a Single Particle Model (SPM) of a battery was developed and discretized. Battery parameter grouping was then performed to reduce the number of parameters required. Using a set of parameters previously identified from a real battery as a benchmark, we generated fitting and validation datasets to assess the methods' runtime and accuracy. The comparative analysis reveals that PSO outperforms the other methods in terms of accuracy and stability, making it highly effective for parameter identification when there is no prior knowledge of the battery's internal parameters. In contrast, LS is better suited for minor adjustments in parameters, particularly for aging batteries, whereas GA lags behind in both computational efficiency and optimality with respect to PSO.
\end{abstract}


\section{Introduction}

As lithium-ion batteries are increasingly used in transportation, energy storage, and consumer electronics, ensuring battery safety and maximizing their potential has become a hot topic of research \cite{b1}. Unfortunately, the equivalent circuit battery model often falls short in fulfilling these tasks. To overcome these limitations, it becomes imperative to model the internal reaction mechanisms of batteries, particularly through electrochemical battery models. The advantage of electrochemical models is that they can effectively reflect the internal information of the battery and offer higher accuracy. However, some of the parameters in electrochemical models cannot be directly measured or are difficult to measure experimentally, and there are many parameters, which makes the model often to be considered  as over-parameterized. This means that two different sets of parameters might yield the same results, posing a significant challenge for the parameter identification of electrochemical models\cite{b2}.

The electrochemical model of batteries generally refers to the P2D model and its variants. The P2D model was originally proposed by Doyle, Fuller, and Newman, hence it is also called the DFN model\cite{b3}. Due to the slow solving speed of the P2D model, there are many simplified models developed for practical applications, such as the Single Particle Model (SPM)\cite{b4}. The SPM model uses a single particle to represent multiple particles in the electrode and ignores changes in the concentration within the electrolyte. This approach speeds up the computation of the model and reduces the number of model parameters. Even a simplified SPM contains many parameters that need to be estimated.

The most widely used methods that have been applied to the parameter identification of electrochemical models include: Least Squares Method (LS)\cite{b5,b6}, Particle Swarm Optimization (PSO)\cite{b7,b8,b9}, and Genetic Algorithm (GA)\cite{b10,b11,b12,b13}. The LS algorithm is an optimization algorithm for solving least squares problems. Its advantages include the ability to converge quickly to local minima and possessing good global search capabilities. The Levenberg-Marquardt method is the most commonly used LS. However, due to its inability to specify bounds on the estimated parameters, it is not suitable for direct use in the parameter estimation of electrochemical models. Researchers typically employ the Trust Region Reflective method to address least squares problems with boundary constraints. PSO is a group-based evolutionary computation technique. The PSO algorithm is simple and easy to implement, does not require complex parameter tuning, and is applicable to a variety of optimization problems. GA is a search algorithm that simulates natural selection and genetic principles, and is a type of evolutionary algorithm. GA does not depend on the specific form of the problem, exhibits strong robustness, and is particularly suitable for handling complex optimization problems with many parameters and numerous local minima.  However, currently, there are no comprehensive studies that provide a horizontal comparison of these three methods in terms of efficiency and optimality when applied to battery parameter identification. Such a comparison would offer valuable guidance for parameter identification in electrochemical models.

This study will first establish a SPM, followed by discretizing the SPM to facilitate model computations. Subsequently, the battery model parameters will be grouped to reduce the number of parameters that need to be estimated. Using this SPM, battery current and voltage data will be generated to validate three parameter identification methods. Finally, a comparative analysis of these three methods will be conducted.

\section{Electrochemical Battery Model}

This section will first introduce the SPM, then proceed to discretize it, and finally group the battery model parameters and select the parameters that need to be identified.

\subsection{The SPM }

The diffusion of lithium in the solid phase of the electrode is described by Fick's second law given by the following Partial Differential Equation (PDE):

\begin{equation}
\frac{\partial c_{s,i}}{\partial t} = \frac{D_{s,i}}{R_{s,i}^2} \frac{\partial}{\partial r} \left( r^2 \frac{\partial c_{s,i}}{\partial r} \right)\label{eq:1}
\end{equation}
where $c_{s,i}$ is the concentration of lithium in the solid phase, $D_{s,i}$ is the solid phase diffusion coefficient, where $i \in \{p,n\}$ with $p$ and $n$ as subscripts for positive and negative electrode,  $R_s$ is the particle radius, $r$ is the radial position within the electrode particle, and $t$ is time.

The boundary conditions for PDE \eqref{eq:1} are given at the center of the electrode particle (\( r = 0 \)) by:

\begin{equation}
\frac{\partial c_{s,i}}{\partial r} = 0\label{eq:2}
\end{equation}
and at the particle's surface $r = R_s$ by:

\begin{equation}
D_{s,i} \frac{\partial c_{s,i}}{\partial r} = \frac{-j_i}{F}  \label{eq:3}
\end{equation}
where $j_i$ is the flux of lithium ions into the particle, and $F$ is Faraday's constant.

The overpotentials are described by the Butler-Volmer equation and given by:
\begin{equation}
\eta_{i} = \frac{2RT}{ F} \sinh^{-1} \left( \frac{- I}{2S_ij_{0,i}} \right) \label{eq:6}
\end{equation}
where $j{0,i}$ is the exchange current density, $\eta_{i}$ is the overpotential, $I$ is the current, $R$ is the gas constant, and $T$ is the temperature. The variable $j_0$ is given by:

\begin{equation}
j_{0,i} = r_{eef,i}  \sqrt{c_e\ c_{ss,i}( c_{ max,i} - c_{ss,i})}  \label{eq:5}
\end{equation}
where $r_{eef,i} $ is the electrode reaction rate, $c_e$ is the lithium-ion concentration in electrolyte, $c_{ss,i}$ is lithium-ion concentration on electrode surface, $c_{\rm max,i}$ is the maximum lithium-ion concentration of electrode.

$S_i$ is defined as:

\begin{equation}
S_i = \frac{3a_iL_i \varepsilon_{s,i}}{R_{s,i}} \label{eq:8}
\end{equation}
with $a_i$, $L_i$ as the surface area and thickness of the negative and positive electrodes, and $\varepsilon_{s,i}$ as the electrode active material volume fraction of the negative and positive electrodes.

The cell voltage is calculated as:
\begin{equation}
V_{bat} = OCP_n(\theta_n)-OCP_p(\theta_p) - (\eta_{p} -\eta_{n} ) -IR_0  \label{eq:10}
\end{equation}
where $\theta_p$ and $\theta_n$ are defined as:

\begin{equation}
\theta_i=\frac{c_{ss,i}}{c_{max,i}}\label{eq:11}
\end{equation}
while $OCP_i(\theta_i)$ is the solid phase potentials of the positive and negative electrodes, respectively, and $R_0$ is the internal resistance.

\subsection{Model Discretization}

This study uses a parabolic equation to approximate certain internal physical processes of the battery, particularly those related to solid phase diffusion in \eqref{eq:1} \cite{b13}. Such spatial discretization turns the PDE into the following set of Ordinary Differential Equations 
\begin{equation}
\frac{d}{dt} \overline{{c}}_{s}=-3\frac{j}{R_{s}}\label{eq:13}
\end{equation}

\begin{equation}
\frac{d}{dt} \overline{c}_{fs}+30\frac{D_{s}}{R_{s}^{2}} \overline{c}_{fs}+\frac{45j}{2R_{s}^{2}}=0 \label{eq:14}
\end{equation}

\begin{equation}
c_{ss}=\overline{c}_{s}+\frac{8R_{s}}{35}\overline{c}_{fs}-\frac{R_{s}}{35D_{s}}j \label{eq:15}
\end{equation}
where $\overline{c}_{fs}$ is the average concentration flux of lithium in active material.

The state-space representation of the  SPM discretized in time with control input $I$ is:
\begin{align}
\begin{bmatrix}
\overline{c}_{s,n}^{(k+1)} \\
\overline{c}_{s,p}^{(k+1)} \\
\overline{c}_{fs,n}^{(k+1)} \\
\overline{c}_{fs,p}^{(k+1)}
\end{bmatrix}
&\!\!=\!\! \begin{bmatrix}
1 &\!\! 0 &\!\! 0 &\!\! 0 \\
0 &\!\! 1 &\!\! 0 &\!\! 0 \\
0 &\!\! 0 &\!\! 1 - \displaystyle \frac{30D_{s,n}\Delta t}{R_{s,n}^2} &\!\! 0 \\
0 &\!\! 0 &\!\! 0 &\!\! 1 - \displaystyle \frac{30D_{s,p}\Delta t}{R_{s,p}^2}
\end{bmatrix} 
\!\!
\begin{bmatrix}
\overline{c}_{s,n}^{(k)} \\
\overline{c}_{s,p}^{(k)} \\
\overline{c}_{fs,n}^{(k)} \\
\overline{c}_{fs,p}^{(k)}
\end{bmatrix} \notag \\
&\quad + \begin{bmatrix}
\displaystyle \frac{\Delta t}{F a_n L_n \varepsilon_{s,n}}  \\
\displaystyle -\frac{ \Delta t}{ F a_p L_p \varepsilon_{s,p}} \\
\displaystyle \frac{15 \Delta t}{2 F R_{s,n} a_n L_n \varepsilon_{s,n}} \\
\displaystyle -\frac{15 \Delta t}{2 F R_{s,p} a_p L_p \varepsilon_{s,p}}
\end{bmatrix} I^{(k)}  \label{eq:17}
\end{align}
where the output equation representing the solid-phase surface concentrations are given by:

\begin{equation}
c_{ss,n}^{(k+1)}=\overline{c}_{s,n}^{(k+1)}+\frac{8R_{s,n}}{35}\overline{c}_{fs,n}^{(k+1)}+\frac{R_{s,n}}{35D_{s,n}} \frac{R_{s,n}}{3Fa_nL_n\varepsilon_{s,n}}I^{(k)}
\label{eq:18}
\end{equation}
\begin{equation}
c_{ss,p}^{(k+1)}=\overline{c}_{s,p}^{(k+1)}+\frac{8R_{s,p}}{35}\overline{c}_{fs,p}^{(k+1)}-\frac{R_{s,p}}{35D_{s,p}}\frac{R_{s,p}}{3Fa_pL_p\varepsilon_{s,p}}I^{(k)}
\label{eq:19}
\end{equation}
where   $\Delta t$ is the sampling time, $(k)$ is the step number.

\subsection{Model Parameter Grouping}

The parameter grouping involves categorizing some parameters with redundant effects and only estimating those that have an impact on the overall computation. For example, in \eqref{eq:1}, the PDE has a common factor of $D_{s,i}/R_{s,i}^2$. Therefore, we do not need to estimate both parameters within this coefficient simultaneously; instead, we choose to estimate \(D_{s,i}\) and set \(R_{s,i}\) to a fixed value. Similarly, the parameters \(a_i\), \(L_i\), and \(\varepsilon_{s,i}\) also appear together in the equation, hence we choose to estimate only \(\varepsilon_{s,i}\). By adopting this approach, we reduced six parameters (\(R_{s,i}\),\(a_i\),\(L_i\)) to be estimated. Thus, the model requires the estimation of 11 parameters in total, as detailed in Table~\ref{tab:1}.

\begin{table}[htbp]
\centering
\caption{List of Parameters to Estimate}
\begin{tabular}{cc}
\hline
\textbf{Variable} & \textbf{Parameter definition} \\ \hline
\(c_{n0}\)  & Initial concentration in negative electrode [${\rm mol}/{\rm m}^3$]  \\ 
\(c_{p0}\) & Initial concentration in positive electrode [${\rm mol}/{\rm m}^3$] \\ 
\(r_{eef,n}\)  & Negative electrode reaction rate [-] \\ 
\(r_{eef,p}\) & Positive electrode reaction rate  [-]\\ 
\(\varepsilon_{s,n}\) & Negative electrode active material volume fraction [-] \\ 
\(\varepsilon_{s,p}\) & Positive electrode active material volume fraction [-] \\ 
\(D_{s,n}\) & Negative electrode diffusivity [${\rm m}^2/{\rm s}$] \\ 
\(D_{s,p}\) & Positive electrode diffusivity [${\rm m}^2/{\rm s}$] \\ 
\(R_{0}\)  & Internal resistance [$\Omega$] \\ 
\(c_{\rm max,n}\) & Maximum concentration in negative electrode [${\rm mol}/{\rm m}^3$]\\ 
\(c_{\rm max,p}\) & Maximum concentration in positive electrode [${\rm mol}/{\rm m}^3$]\\ \hline
\end{tabular}
\label{tab:1}
\end{table}

\section{Experimental Design}

This section introduces experimental design, which specifically includes three steps: dataset generation, simulation environment and optimization settings for the optimizers, results validation.

\subsection{Dataset Generation}

The SPM introduced in Section 2, along with a set of pre-estimated battery parameters from a real battery, is employed to generate a series of battery current and voltage dataset. This dataset includes discharge rates of 0.1C, 0.2C, 0.3C, 0.4C, 0.5C, 0.6C, 0.7C, 0.8C, 0.9C, 1C, and a Dynamic Stress Test (DST). Battery current and voltage data at a discharge rate of 0.5C are used for the identification of battery parameters, which is referred to as the fitting dataset. All other data are employed to validate the results of the battery parameter identification and are referred to as the validation dataset. Parameter estimation algorithms may only find local optima, which could lead to low voltage errors at a 0.5C discharge rate. However, these same settings may result in significant errors under other operating conditions. Therefore, using a validation set is crucial to ensure whether the algorithms have indeed identified a global optimum solution.

\subsection{Simulation Environment and Optimization Settings}

The three methods, LS, PSO, and GA, all require the boundaries for the parameters to be identified. Therefore, the pre-estimated battery parameters introduced in the previous section are used as reference values for true parameters. Based on these reference values, the upper boundary for the parameters is set at 1.5 times the reference values, and the lower boundary is set at 0.5 times the reference values. This approach establishes the boundaries for all 11 parameters that need to be estimated.

PSO and GA do not require the setting of initial battery parameter values; however, the LS algorithm does. Therefore, for LS, within the boundary constraints of each parameter, one hundred sets of initial battery parameters are randomly generated using a uniform distribution to serve as the initial battery parameters for the LS algorithm. The bounds of the initial battery parameters are as previously mentioned.

All programs in this study are written in Python and the simulations are conducted on a computer with the following specifications: Windows 10 Enterprise, Intel(R) Core(TM) i5-1145G7 processor and 16 GB of RAM.

The implementation of the LS algorithm is performed using the \texttt{optimize.curve\_fit} function from the \texttt{scipy} library (version 1.11.4). This function defaults to the Trust Region Reflective method, incorporating the parameter boundaries and initial battery parameter values introduced in the previous section as optimization inputs, with all other parameters set to their default values.

The implementation of the PSO algorithm is conducted using the \texttt{pso} function from the \texttt{pyswarm} library (version 0.6). In addition to the parameter boundaries provided as input to the algorithm, other parameter settings are configured as follows: the maximum number of iterations \texttt{maxiter} is set to 100, and the minimum function tolerance \texttt{minfunc} is established at \(10^{-8}\). These settings ensure a balance between computational efficiency and convergence accuracy, facilitating the algorithm's application in complex optimization problems.

The implementation of the GA is carried out using the \texttt{GA} function provided by the \texttt{pygad} library (version 3.3.1). Similar parameter boundaries are employed as in other algorithmic implementations. For other parameters include \texttt{num\_generations}, set to 300, which determines the number of cycles the algorithm will run to evolve the solutions. The \texttt{num\_parents\_mating} is set to 4, indicating the number of parents selected for breeding in each generation. Furthermore, the population size \texttt{sol\_per\_pop} is configured at 50, ensuring a diverse genetic pool, and \texttt{num\_genes} per individual solution is defined as 11, corresponding to the 11 parameters to be estimated.

\subsection{Results Validation}

We use the dataset at a 0.5C discharge rate for parameter identification via three methods to obtain optimized battery parameters. These optimized battery parameters are then used to rerun the SPM to calculate voltage data under the validation dataset. Subsequently, we calculate the error between the voltages by the battery model using the optimized parameters and those using the reference parameters. This error reflects the accuracy of the parameter estimation by the three parameter identification methods. 

For LS, given the necessity to set initial battery parameter values, we use the 100 sets of initial batttery parameters generated in the previous step as the starting points. We run the LS algorithm 100 times to obtain 100 sets of optimized parameters. For PSO and GA, which do not require initial values, we exclude the effects of randomness by running each algorithm 20 times.

\section{Results}

This section will present the results of using the LS, PSO, and GA algorithms for parameter identification. The comparative results of the three algorithms are presented in Table~\ref{tab:2} in terms of running time and error.

\begin{table*}[htbp]

\centering
\caption{Comparison of Algorithm Performance}

\begin{tabular}{c|ccc}
\hline
\textbf{Method} & \textbf{Runtime (SD) [min]} & \textbf{Fitting RMSE (SD) [mV]} & \textbf{Validation RMSE (SD) [mV]} \\
\hline
TRF&2.907 (4.715) &225.6 (186.5) &226.5 (181.5) \\
PSO&39.99 (12.32) &5.534 (4.938) &14.27 (4.492)\\
GA&67.94 (9.448) &12.37 (4.855) &31.35 (11.98)\\
\hline
\end{tabular}
\label{tab:2}
\end{table*}

\subsection{LS}

\begin{figure}[htbp]
\centerline{\includegraphics[width=0.5\textwidth]{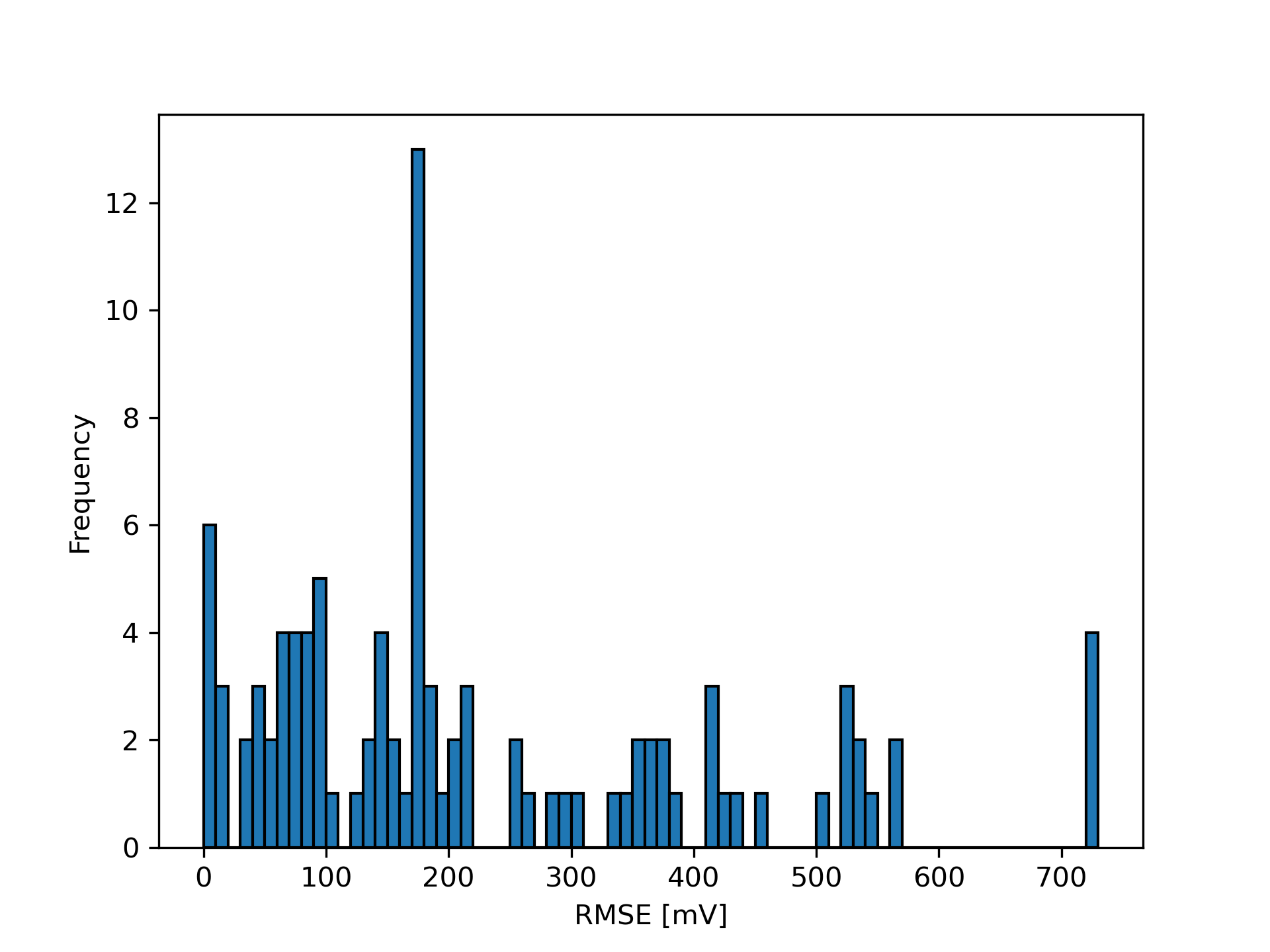}}
\caption{ Distribution of RMSE for the TRR algorithm on the fitting dataset}
\label{fig:1}
\end{figure}

\begin{figure}[htbp]
\centerline{\includegraphics[width=0.5\textwidth]{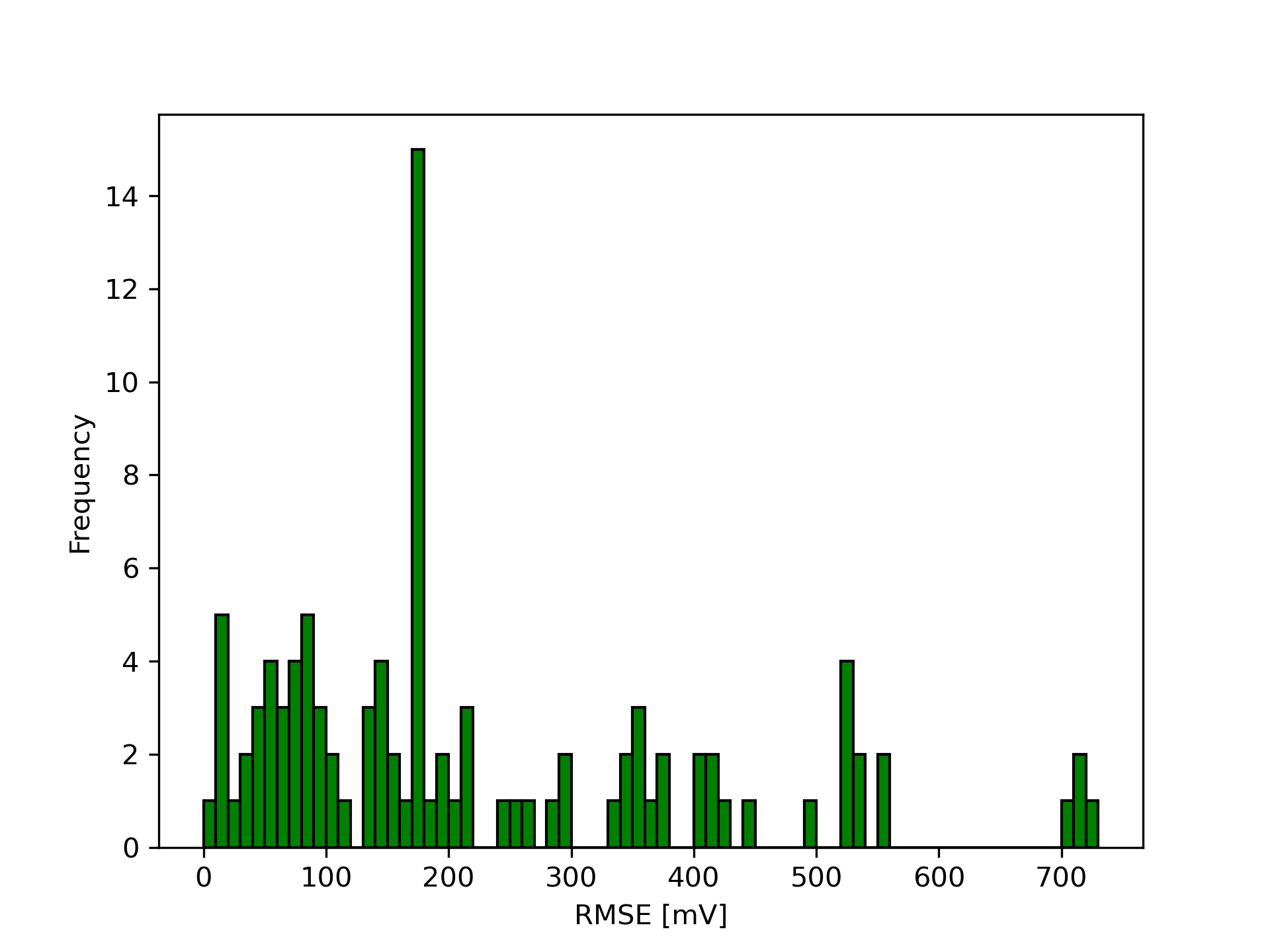}}
\caption{ Distribution of RMSE for the TRR algorithm on the validation dataset}
\label{fig:2}
\end{figure}

The average execution time for parameter identification using the LS algorithm is 2.907 min\, with a standard deviation (SD) of 4.715 min. The fastest run time recorded is 7.488 s, and the longest is 17.76 min. This indicates that the computational time distribution for the LS method is quite dispersed, heavily dependent on the choice of initial battery parameter values.
fitting dataset is shown in Fig.~\ref{fig:1}, while the RMSE distribution on the validation dataset is presented in Fig.~\ref{fig:2}. Frequency are statistically tallied at intervals of 10 mV. The LS algorithm has an RMSE of 225.6 mV (SD:186.5 mV) on the fitting dataset. On the validation dataset, the average RMSE is 226.5 mV (SD:181.5 mV).

The Fig.~\ref{fig:1} shows that the LS algorithm, when estimating parameters, shows a widely scattered error distribution. Notably, the RMSE exceeds 500 mV for 13 optimized parameter sets on the fitting dataset, indicating that LS sometimes fails to find the global optimum. On the fitting dataset, six optimized parameter sets have an error below 10 mV; of these, one also shows below 10 mV on the validation dataset, while the others remain under 20 mV. This indicates that while the LS algorithm can identify global optima, its success heavily depends on the choice of initial battery parameter values.

\subsection{PSO}
The average execution time for parameter identification using the PSO method is 39.99 min (SD:12.33 min). The fastest run time recorded is 22.73 min, and the longest is 63.58 min. 

\begin{figure}[htbp]
\centerline{\includegraphics[width=0.5\textwidth]{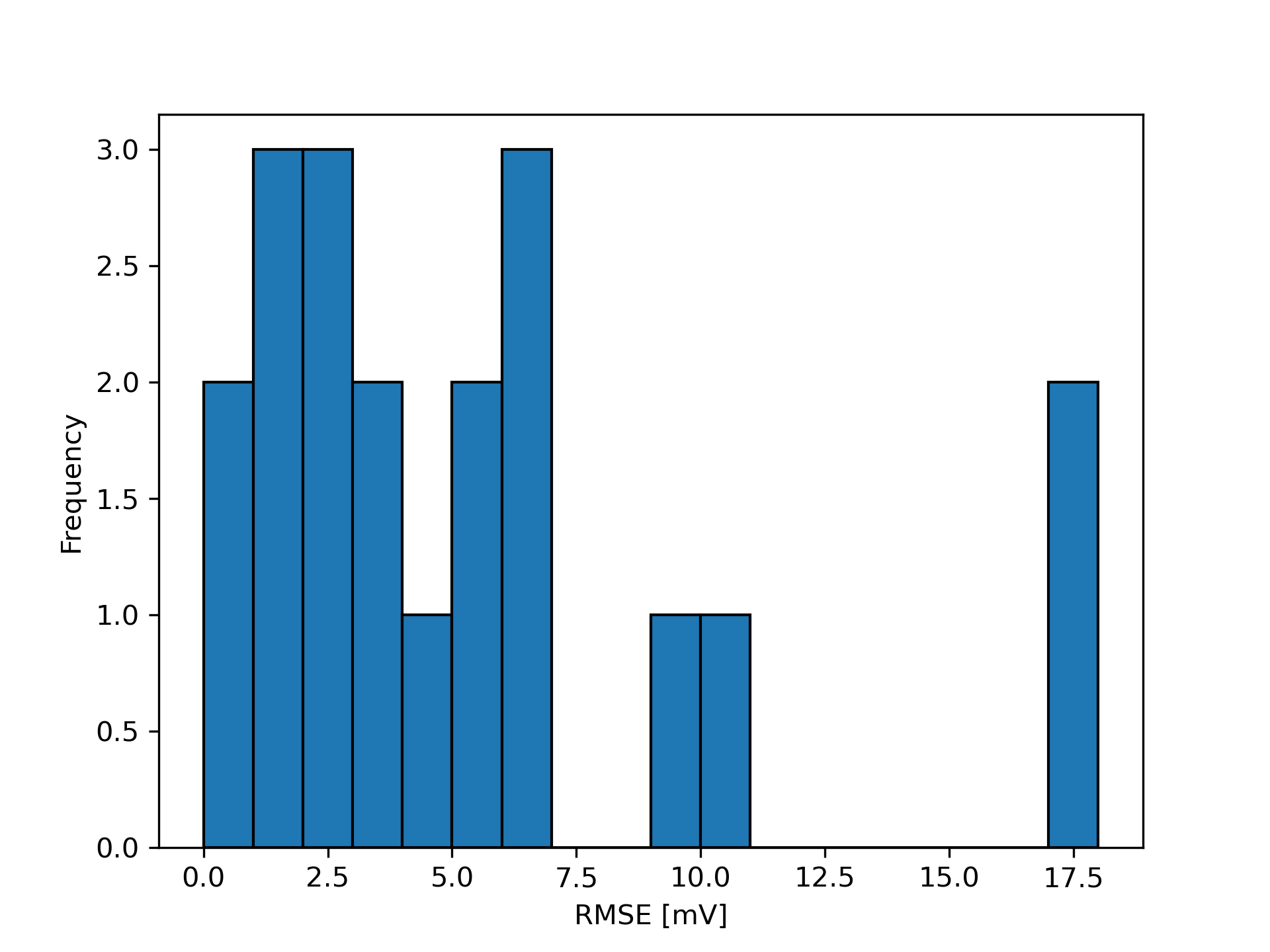}}
\caption{ Distribution of RMSE for the PSO algorithm on the fitting dataset}
\label{fig:3}
\end{figure}

\begin{figure}[htbp]
\centerline{\includegraphics[width=0.5\textwidth]{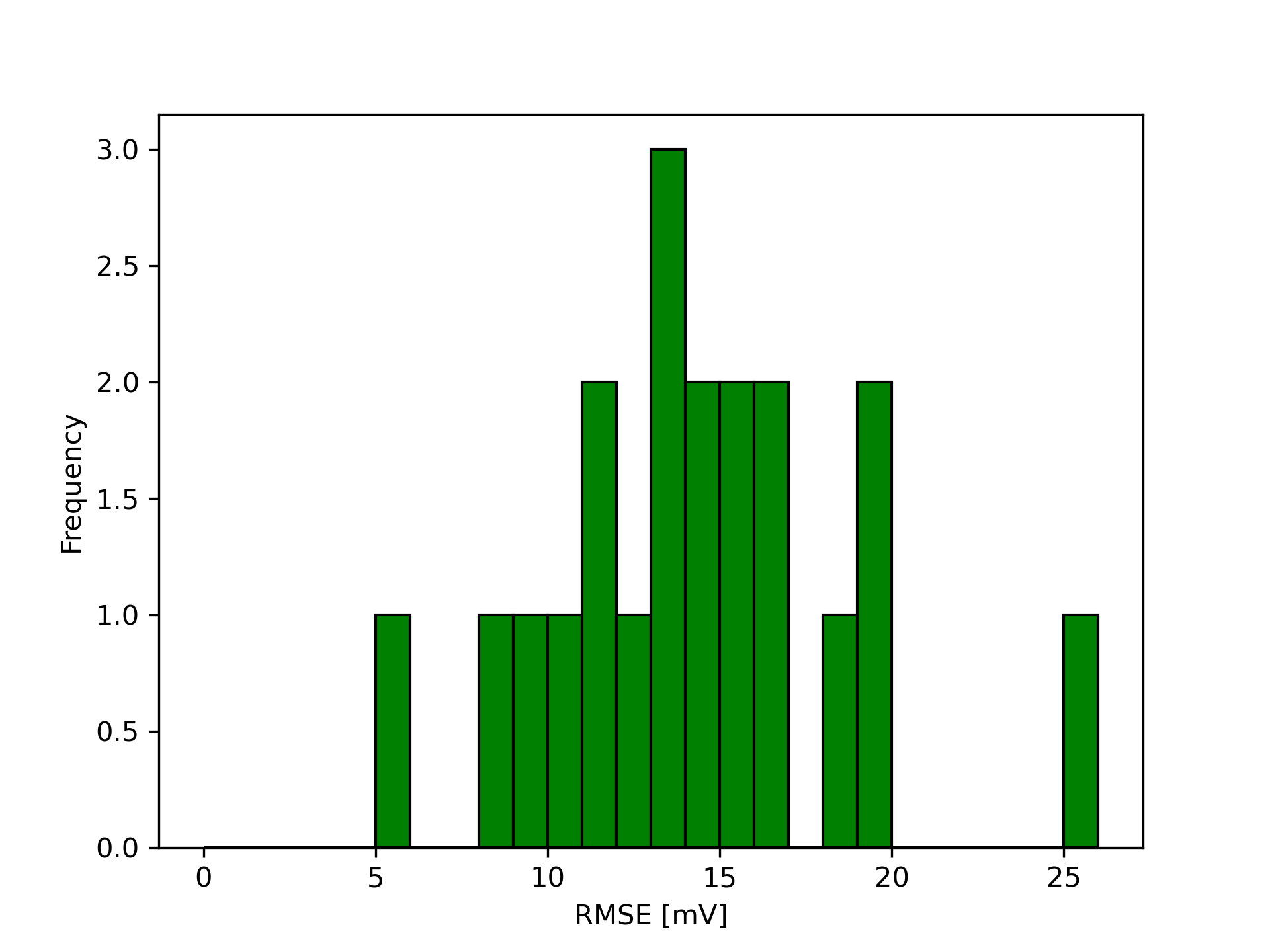}}
\caption{ Distribution of RMSE for the PSO algorithm on the validation dataset}
\label{fig:4}
\end{figure}

The distribution of RMSE for the PSO algorithm on the fitting dataset is shown in Fig.~\ref{fig:3}, while the RMSE distribution on the validation dataset is presented in Fig.~\ref{fig:4}.Frequency are statistically tallied at intervals of 1 mV. The PSO algorithm has an RMSE of 5.534 mV (SD: 4.938 mV) on the fitting dataset. On the validation dataset, the average RMSE is 14.27 mV (SD:4.492 mV).

From Fig.~\ref{fig:3}, it is evident that all 20 optimized parameter sets achieve RMSE of less than 20 mV on the fitting dataset, with 17 optimized parameter sets below 10 mV and only three optimized parameter sets greater than 10 mV. Fig.~\ref{fig:4} shows that, in the validation dataset, only one optimized parameter set exhibits an RMSE greater than 20 mV. These results indicate that the PSO algorithm has high accuracy and consistently identifies global optima with stability.

\subsection{GA}
The average execution time for parameter identification using the GA algorithm is 67.94 min (SD:9.448 minutes). The fastest run time recorded is 51.26 min, and the longest is 83.86 min. 

\begin{figure}[htbp]
\centerline{\includegraphics[width=0.5\textwidth]{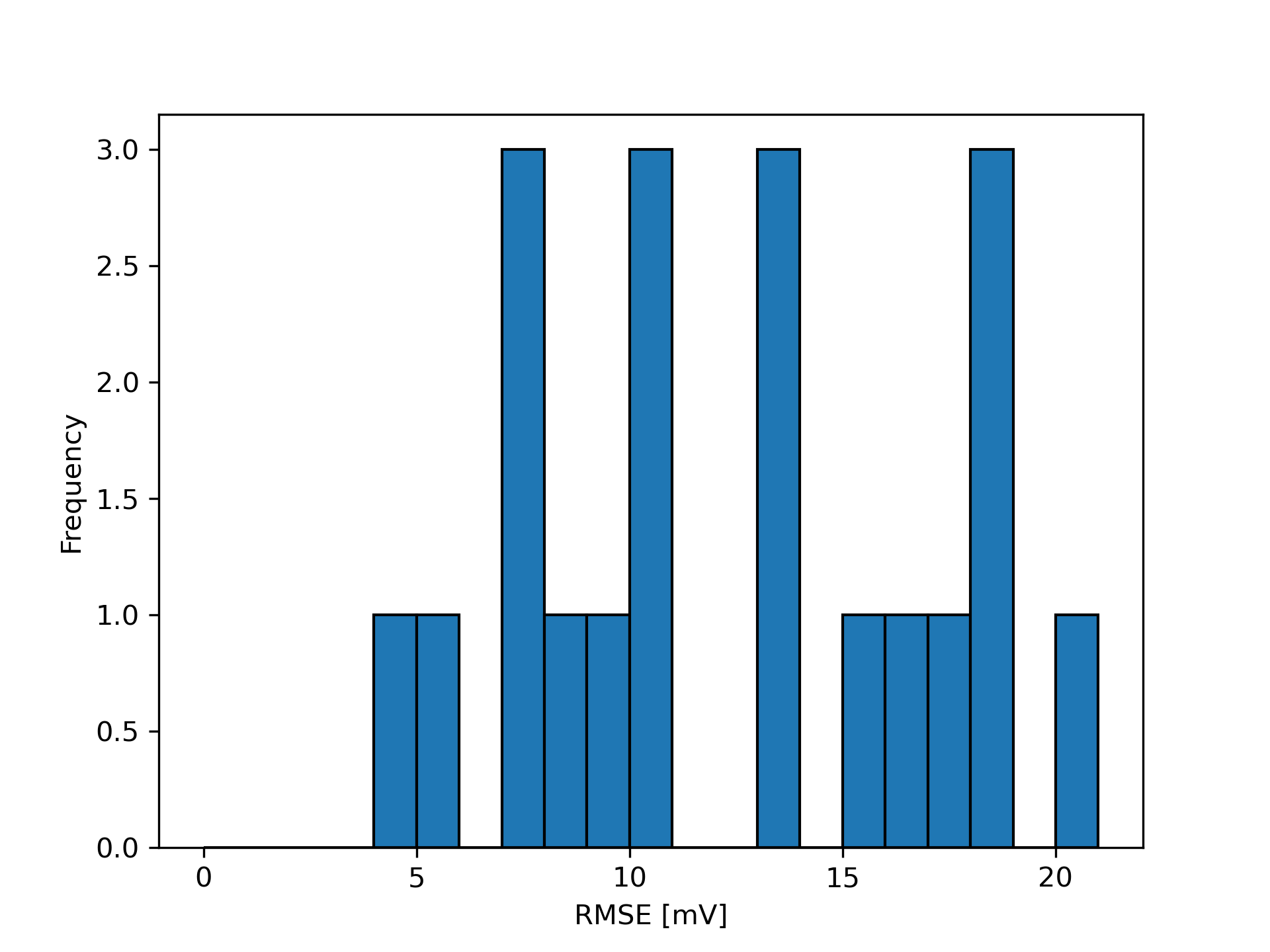}}
\caption{ Distribution of RMSE for the GA algorithm on the fitting dataset}
\label{fig:5}
\end{figure}

\begin{figure}[htbp]
\centerline{\includegraphics[width=0.5\textwidth]{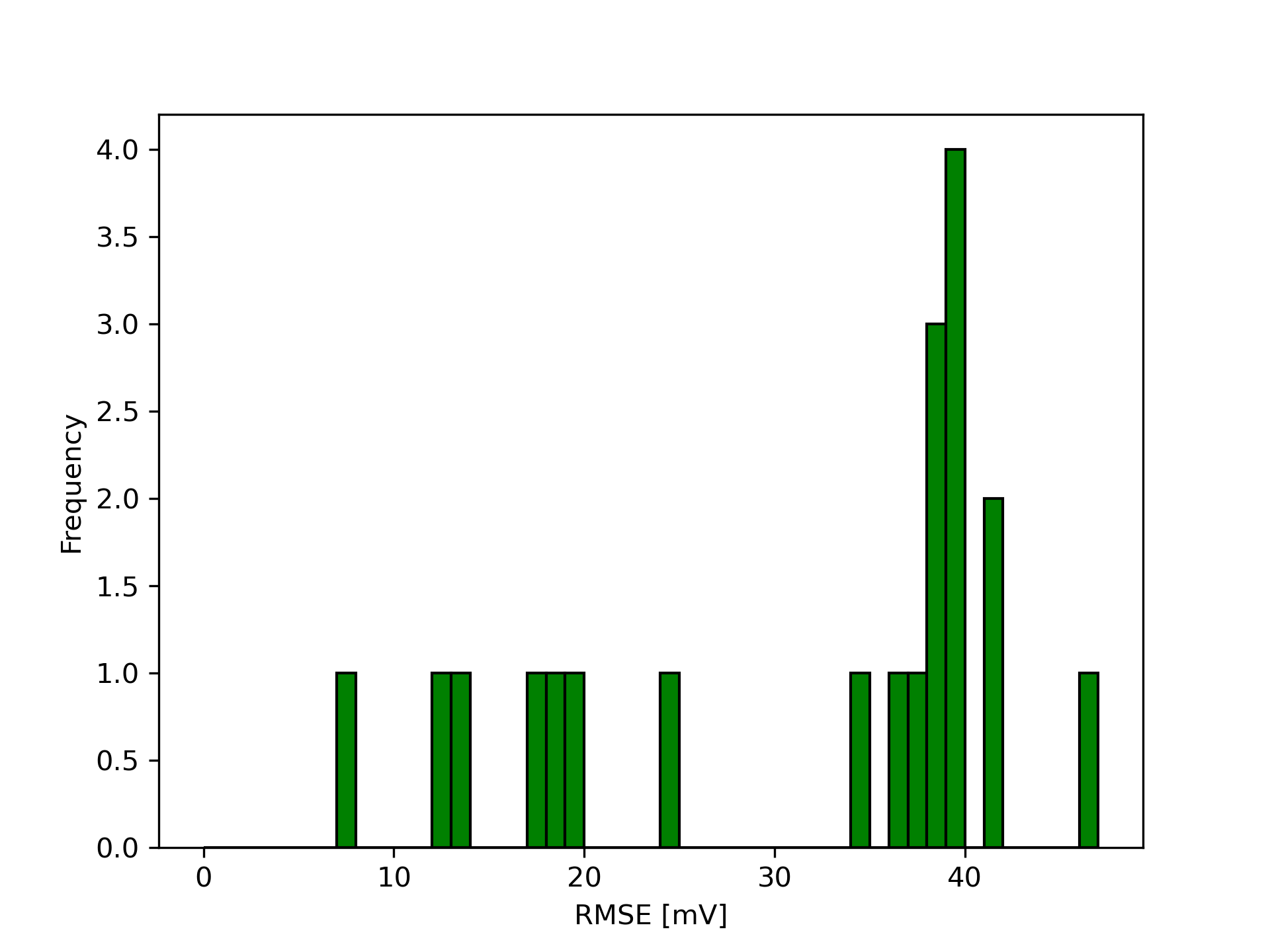}}
\caption{ Distribution of RMSE for the GA algorithm on the validation dataset}
\label{fig:6}
\end{figure}

The distribution of RMSE for the GA algorithm on the fitting dataset is shown in Fig.~\ref{fig:5}, while the RMSE distribution on the validation dataset is presented in Fig.~\ref{fig:6}. Frequency are statistically tallied at intervals of 1 mV. The GA algorithm has an RMSE of 12.37 mV (SD: 4.855 mV) on the fitting dataset. On the validation dataset, the average RMSE is 31.35 mV (SD:11.98 mV).

Fig.~\ref{fig:5} illustrates that, among the 20 optimized parameter sets evaluated using the GA on the fitting dataset, only one optimized parameter sets greater than 20 mV while the remaining 19 are below 20mV, with seven optimized sets under 10 mV. According to Fig.~\ref{fig:6}, in the validation parameter dataset, all optimized parameter sets remain below 50 mV, and six of these are under 20 mV.

\section{Discussions}

From the perspective of algorithmic efficiency, the results from Table~\ref{tab:2} indicate that the LS algorithm has the fastest average execution time of 2.907 min. However, it shows a high variance in performance, with durations extending up to 17.76 min. This variability is attributed to the LS algorithm's sensitivity to initial battery parameter settings. In scenarios where there is limited knowledge about the internal parameters of the battery, it becomes challenging to narrow the parameter boundaries and provide appropriate initial estimates, resulting in reduced stability. Compared to the LS algorithm, both the PSO and GA take more time to compute, with PSO being  faster than GA. However, it is important to highlight that the runtime of these algorithms depends on the initial settings like maximum number of iterations, as discussed in \mbox{Section 3}. These settings greatly affect how well and how quickly the algorithms can estimate parameters. Despite this, under the settings used in this paper, PSO achieves higher estimation accuracy compared to GA, while requiring less time.

From the perspective of algorithmic optimality, the PSO algorithm achieves the best results, obtaining the lowest RMSE and high stability. The LS algorithm also yields satisfactory results, with six optimized parameter sets in the fitting dataset below 10 mV and in the validation dataset below 20 mV. However, compared to its 100 optimized parameter sets, the probability of obtaining optimal parameters is only 6\%. Nevertheless, if there is some information of battery parameters, reducing the initial estimation range or fine-tuning around parameters of aged batteries makes LS a good choice due to its faster computation speed. The GA method does not perform as well as PSO in terms of speed and accuracy. For a completely new battery with unknown internal parameters information, the PSO algorithm may be the best choice among the methods compared in this study.

\section{Conclusion}

This paper compares the efficiency and optimality of three common parameter identification methods for electrochemical battery models: LS, PSO, and GA. Initially, a SPM was established, followed by the discretization of this model. The number of parameters needing identification was then reduced by grouping the battery parameters. A set of battery parameters, previously identified based on a real battery, was used as a reference. These parameters, alongside the SPM, generated fitting and validation datasets used to compare the runtime and accuracy of the three methods. The results indicate that PSO offers the best accuracy and stability, making it suitable for identifying battery parameters when there is no prior knowledge of the battery parameters. LS is more suited for fine-tuning parameters within a narrow range, such as for aging batteries, while GA is inferior to PSO in both computational time and accuracy.

In future research, there is a need to focus more on battery parameter grouping methods to effectively reduce the number of parameters that need to be estimated. At the same time, sensitivity analysis for battery parameter estimation is crucial. This helps us identify the most sensitive parameters, allowing us to further narrow their estimated range and speed up the battery parameter estimation process.

\end{document}